\documentclass[11pt]{article}
\usepackage{graphicx}
\newcommand{\be}{\begin{equation}}
\newcommand{\ee}{\end{equation}}
\newcommand{\bea}{\begin{eqnarray}}
\newcommand{\eea}{\end{eqnarray}}

\begin{document}
\vspace{0.5in}
\begin{center}
{\LARGE{\bf Logarithmic Potential with Super-Super- Exponential Kink Profiles and Tails}}
\end{center} 

\vspace{0.2in} 

\begin{center}
{{\bf Avinash Khare}} \\
{Physics Department, Savitribai Phule Pune University, \\
 Pune 411007, India}
\end{center}

\begin{center} 
{{\bf Avadh Saxena}} \\
{Theoretical Division and Center for Nonlinear Studies, 
Los Alamos National Laboratory, Los Alamos, New Mexico 87545, USA}
\end{center}

\vspace{0.9in}
\noindent{\bf {Abstract}}
We consider a novel one dimensional model of a logarithmic potential which 
has super-super-exponential kink profiles as well as kink tails.  We provide 
analytic kink solutions of the model -- it has 3 kinks, 3 mirror kinks and the 
corresponding antikinks. While some of the kink tails are super-super-exponential, 
some others are super-exponential whereas the remaining ones are exponential. 
The linear stability analysis reveals that there is a gap between the zero mode and 
the onset of continuum.   Finally, we compare this potential and its kink solutions 
with those of very high order field theories harboring seven degenerate minima and 
their attendant kink solutions, specifically $\phi^{14}$, $\phi^{16}$ and $\phi^{18}$.

\section{Introduction} 

Recently there has been a growing interest in higher order field theories which admit kink 
solutions with a power law tail at either both the ends or a power law tail at one end and an 
exponential tail at the other end \cite{KCS, Chapter}. An example of the latter is the $\phi^8$ 
potential studied in the context of mesons \cite{Lohe}. This is different from  almost all the 
kink solutions that have been discussed during the last four decades 
where the kink solutions have exponential tail at both the ends \cite{manton1, manton2}.  
The discovery of these power law kinks \cite{Gomes, Bazeia, Guerrero, Mello} has raised 
many interesting questions related to the strength and the range of the kink-kink (KK) and 
kink-antikink (K-AK) force \cite{gani1, manton3}, the possibility of resonances \cite{christov} 
and scattering \cite{gani2}, stability analysis of such kinks \cite{Gomes}, etc.  

Very recently we have introduced a whole family of potentials which exhibits kinks with a power-law 
tail \cite{KS}, a super-exponential tail \cite{pradeep} as well as a power-tower tail \cite{powertower}.  
The potential with super-exponential tail, $V(\phi) = (1/2)(\phi\ln\phi)^2$ \cite{pradeep} arises in 
the context of infinite order phase transitions, whereas higher order field theories can model 
successive (and multiple) first and second order phase transitions \cite{KCS, Chapter}. Thus, 
the logarithmic potential proposed here could have similar physical significance. 

The paper is structured as follows.  In Sec. 2 we introduce a logarithmic  potential 
with super-super-exponential kink tail and explicitly obtain the three kinks,
three mirror kinks and the corresponding six antikink solutions.  We  carry out the 
stability analysis for the different types of kink solutions in Sec. 3 and show that there 
is a gap between the zero mode and the continuum. In Sec. 4 we briefly  
describe the interaction between the different kinks and the antikinks.  In section 5 we 
calculate the kink mass for the three kinks.  In section 6 we compare these results with 
the corresponding ones in specific higher order field theories ($\phi^{14}$, $\phi^{16}$ 
and $\phi^{18}$). Finally in Sec. 7 we summarize the results obtained in this paper and 
point out some of the open problems.

\section{Model and Corresponding Kink Solutions} 

Let us consider the following logarithmic potential
\bea\label{1}
V(\phi) = (1/2) \phi^2 [(1/2)\ln(\phi^2)]^2 ~
\left ([(1/2)\ln[(1/2)\ln(\phi^2)]^2]^2\right)^2\,, 
\eea
which is depicted in Fig. 1. We then find that
\begin{figure}[h] 
\includegraphics[width= 5.1 in]{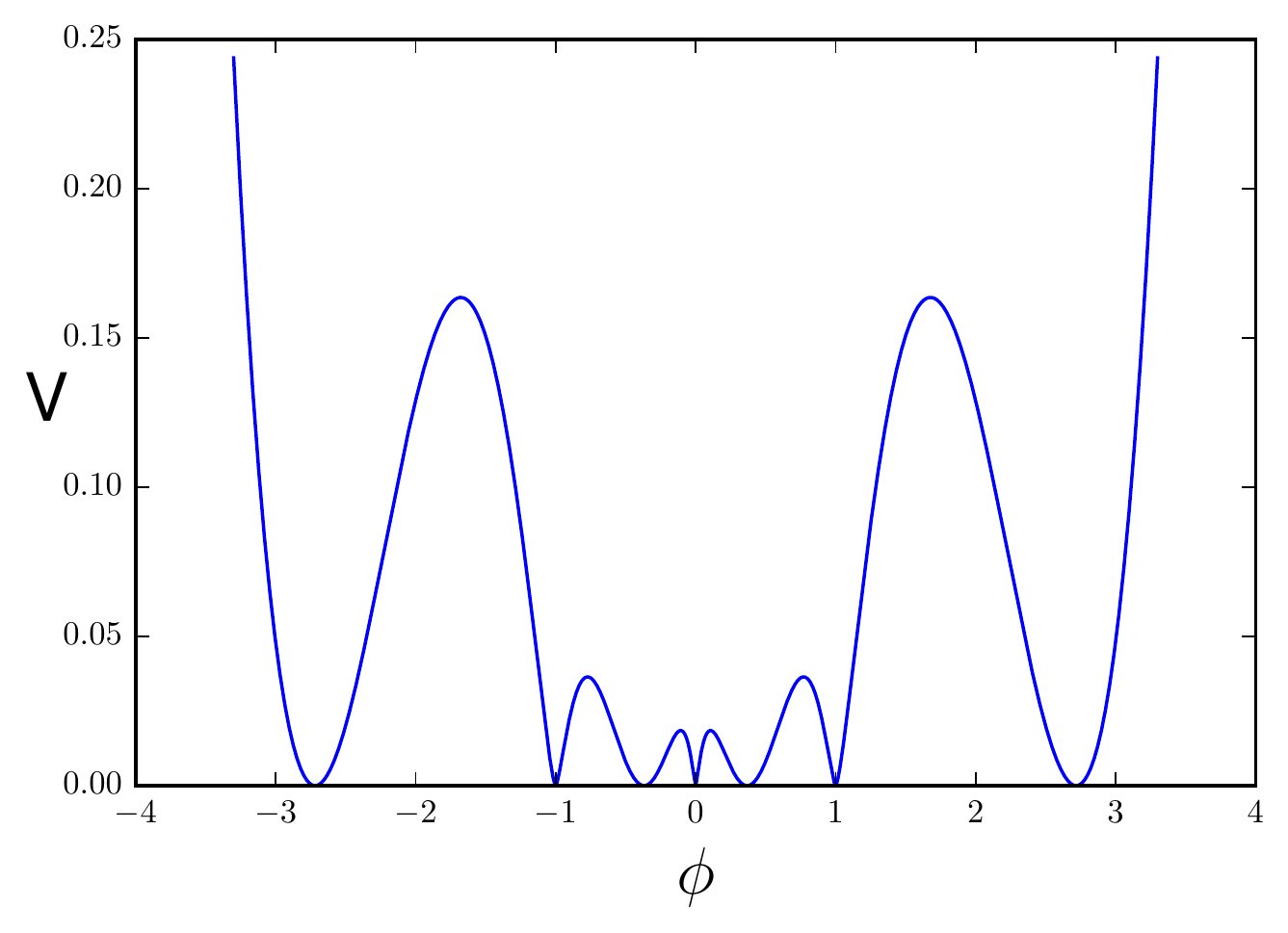}
\caption{Logarithmic potential $V(\phi)$ given by Eq. (1). Note the seven minima at 0,
$\pm1/e$, $\pm1$ and $\pm e$ as well as six symmetric maxima. The potential is smooth 
and there is no cusp either at $\phi=0$ or at $\phi=\pm1$. }
\end{figure} 
\bea\label{2}
\frac{dV}{d\phi} &=& \phi [(1/2)\ln(\phi^2)]
(1/2)\ln[(1/2)\ln(\phi^2)]^2 \nonumber \\ 
&&\times\left ((1/2)\ln[(1/2)\ln(\phi^2)]^2[(1/2)\ln(\phi^2 e^2)+1] \right )\,. 
\eea
This potential has 7 degenerate minima with $V_{min} =0$ at $\phi = 0, \pm 1/e, 
\pm 1, \pm e$ and six maxima (see Fig. 1) which are solutions of the equation:  
\be\label{3}
\left ((1/2)\ln[(1/2)\ln(\phi^2)]^2 [(1/2)\ln(\phi^2 e^2)+1] \right ) = 0\,. 
\ee
It is worth noting that the values of the potential curvature at the seven degenerate minima are
\be\label{3a}
V''(0) = V''(\pm 1) = \infty\,,~~ V''(\pm 1/e) = V''(\pm e) = 1\,.
\ee
Note that the potential is smooth at $\phi=0$ (and at other minima) and there is no cusp there. 
Thus this model will have 3 kink solutions, 3 mirror kink solutions and the corresponding 6 
antikinks. All these kinks and antikinks are solutions of the self-dual equation
\be\label{4}
\frac{d\phi}{dx} = \pm \phi [(1/2)\ln(\phi^2)] 
(1/2)\ln[(1/2)\ln(\phi^2)]^2\,.
\ee 

\subsection{\bf Three Kinks, 3 Mirror Kinks and 6 Anti-kink Solutions}
 
{\bf Solution I}

The kink solution from $0$ to $1/e$ is given by
\be\label{5}
\phi^{I}_{K}(x) = e^{-e^{e^{-x}}}\,,
\ee
and it is easy to check that it is the solution of the self-dual Eq. (\ref{4})
with $+$ve sign. 
In particular,
\be\label{6}
\lim_{x \rightarrow -\infty} \phi(x) = e^{-e^{e^{-x}}}\,,~~
\lim_{x \rightarrow \infty} \phi(x) = \frac{1}{e} - e^{-(x+1)}\,.
\ee
Notice that around $\phi = 0$, the kink tail is super-super-exponential
while around $\phi = /1e$ the tail is exponential.  To our knowledge, this
is the {\it first example} of a kink with a super-super-exponential profile 
and tail. The kink profile is depicted in Fig. 2 (and its magnified version  
is shown in Fig. 3).

On the other hand,  
\be\label{7}
\phi^{I}_{maK}(x) =- e^{-e^{e^{-x}}}\,,
\ee
is the solution of the self-dual Eq. (\ref{4}) with $-$ve sign. 
Note that as $x \rightarrow -\infty$ to 
$+\infty$, $\phi$ goes from $0$ to $-1/e$, i.e. it corresponds to the mirror 
antikink  (maK) associated with the kink solution I, as given by Eq. (\ref{5}).

\vspace{0.2in} 
\noindent{\bf Solution II}

The corresponding mirror kink solution from -$1/e$ to $0$ is given by
\be\label{8}
\phi^{I}_{mK}(x) = - e^{-e^{e^{x}}}\,.
\ee
It is easy to check that it is the solution of the self-dual Eq. (\ref{4})
with $+$ve sign. Note that as $x \rightarrow -\infty$ to 
$+\infty$, $\phi$ goes from $-1/e$ to $0$. 

On the other hand,  
\be\label{9}
\phi^{I}_{aK}(x) = e^{-e^{e^{x}}}\,,
\ee
is the solution of the self-dual Eq. (\ref{4})
with $-$ve sign. Note that as $x \rightarrow -\infty$ to 
$+\infty$, $\phi$ goes from $1/e$ to $0$, i.e. it corresponds to  
the antikink associated with the kink solution (\ref{5}).

It is useful to mention the relationship 
\be\label{10}
\phi^{I}_{maK}(x) = - \phi^{I}_{K}(x)\,,~~ 
\phi^{I}_{aK}(x) = - \phi^{I}_{mK}(x)\,.  
\ee
We will see below that such a relationship also exists for $\phi_K^{II}(x)$ and 
$\phi_K^{III}(x)$.

\begin{figure}[h] 
\includegraphics[width= 5.1 in]{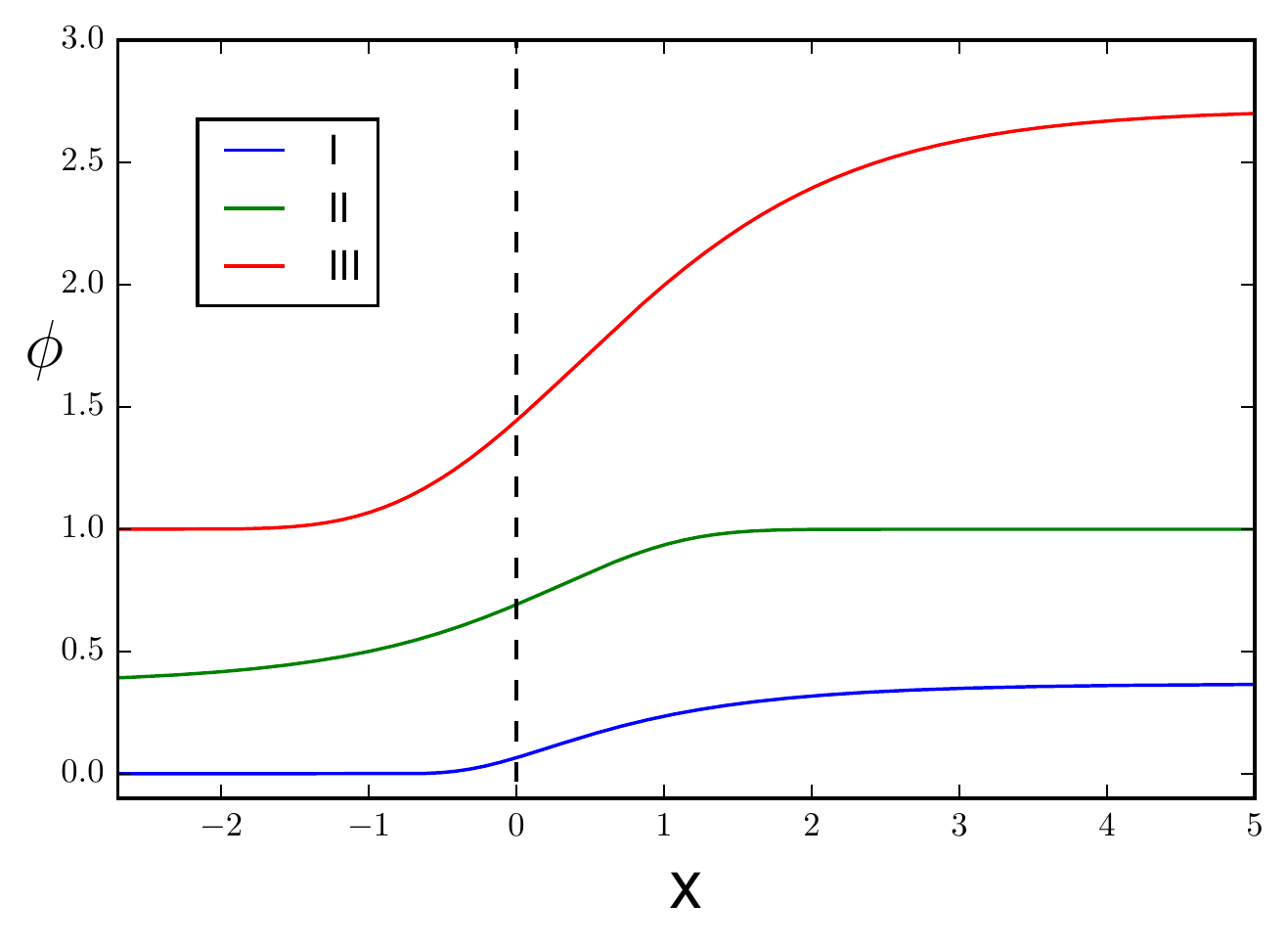}
\caption{The profiles of the three different super-super-exponential kink solutions. 
The kink tails behave as discussed in the text along with the asymptotic values of 
the field $\phi$: 0, $1/e$, 1, $e$. }
\end{figure} 

\vspace{0.2in} 
\noindent{\bf Solution III}

The kink solution from $1/e$ to $1$ is given by
\be\label{11}
\phi^{II}_{K}(x) = e^{-e^{-e^{x}}}\,,
\ee
and it corresponds to the solution of the self-dual Eq. (\ref{4})
with $+$ve sign. 
In particular,
\be\label{12}
\lim_{x \rightarrow -\infty} \phi(x) = \frac{1}{e} + e^{(x-1)}\,,~~
\lim_{x \rightarrow \infty} \phi(x) = 1 - e^{-e^{x}}\,.
\ee
Notice that around $\phi = 1$, the kink tail is super-exponential while it 
is exponential around $\phi = 1/e$.  The kink profile is depicted in Fig. 2  
(and its magnified version is shown in Fig. 4). 

On the other hand,  
\be\label{13}
\phi^{II}_{maK}(x) =- e^{-e^{-e^{x}}}\,,
\ee
is the solution of the self-dual Eq. (\ref{4})
with $-$ve sign. Note that as $x \rightarrow -\infty$ to 
$+\infty$, $\phi$ goes from $-1/e$ to $-1$, i.e. it corresponds to the mirror 
antikink associated with the kink solution II, i.e. Eq. (\ref{11}).

\vspace{0.2in} 
\noindent{\bf Solution IV}

The corresponding mirror kink solution from $-1$ to $-1/e$ is given by
\be\label{14}
\phi^{II}_{mK}(x) = - e^{-e^{e^{x}}}\,.
\ee
It is easy to check that it is the solution of the self-dual Eq. (\ref{4})
with $+$ve sign. Note that as $x \rightarrow -\infty$ to 
$+\infty$, $\phi$ goes from $-1$ to $-1/e$. 

On the other hand,  
\be\label{15}
\phi^{II}_{aK} = e^{-e^{e^{x}}}\,,
\ee
is the solution of the self-dual Eq. (\ref{4})
with $-$ sign. Note that as $x \rightarrow -\infty$ to 
$+\infty$, $\phi$ goes from $1/e$ to $0$, i.e. it corresponds to  
the antikink associated with the kink solution (\ref{11}).

It is worth pointing out that the second kink solution and the corresponding
mirror kink and antikinks as given by Eq. (\ref{11}) and Eqs. (\ref{13}) to
(\ref{15}) also satisfy the relationships in Eq. (\ref{10}).

\vspace{0.2in} 
\noindent{\bf Solution V}

The kink solution from $1$ to $e$ is given by
\be\label{16}
\phi^{III}_{K}(x) = e^{e^{-e^{-x}}}\,, 
\ee
and it corresponds to the solution of the self-dual Eq. (\ref{4})
with $+$ve sign. 
In particular,
\be\label{17}
\lim_{x \rightarrow -\infty} \phi(x) = 1 + e^{-e^{-x}}\,,~~
\lim_{x \rightarrow \infty} \phi(x) = e - e^{-(x-1)}\,.
\ee
Notice that around $\phi = 1$, the kink tail is super-exponential while it 
is exponential around $\phi = e$.  The kink profile is depicted in Fig. 2  
(and its magnified version is shown in Fig. 5). 

On the other hand,  
\be\label{18}
\phi^{III}_{maK}(x) =- e^{e^{-e^{-x}}}\,,
\ee
is the solution of the self-dual Eq. (\ref{4})
with $-$ve sign. Note that as $x \rightarrow -\infty$ to 
$+\infty$, $\phi$ goes from $-1$ to $-e$, i.e. it corresponds to the mirror 
antikink associated with the kink solution III, i.e. Eq. (\ref{16}).

\vspace{0.2in} 
\noindent{\bf Solution VI}

The corresponding mirror kink solution from $-e$ to $-1$ is given by
\be\label{19}
\phi^{III}_{mK}(x) = - e^{e^{-e^{x}}}\,.
\ee
It is easy to check that it is the solution of the self-dual Eq. (\ref{4})
with $-$ve sign. Note that as $x \rightarrow -\infty$ to 
$+\infty$, $\phi$ goes from $-e$ to $-1$. 

On the other hand the mirror kink,  
\be\label{20}
\phi^{III}_{aK} = e^{e^{-e^{x}}}\,,
\ee
is the solution of the self-dual Eq. (\ref{4})
with $+$ve sign. Note that as $x \rightarrow -\infty$ to 
$+\infty$, $\phi$ goes from $e$ to $1$, i.e. it corresponds to the 
antikink associated with the kink solution (\ref{16}).

It is worth pointing out that the third kink solution and the corresponding
mirror kink and antikinks as given by Eq. (\ref{16}) and Eqs. (\ref{18}) to
(\ref{20})  also satisfy the relationships in Eq.  (\ref{10}).

\section{Stability of Kink Solutions}

We now perform the kink stability analysis for all three kink solutions as
given by Eqs. (\ref{5}), (\ref{11}) and (\ref{16}) and show that for all 
of them there is a gap between the zero mode and the onset of the
continuum.  

The kink potential $V_K(x)$ which appears 
in the stability equation
\be\label{2.1}
-\frac{d^2 \psi(x)}{dx^2} + V_K(x) \psi(x) = \omega^2 \psi(x)\,,
\ee
can be calculated using the relationship 
$V_K(x) = \frac{d^2 V(\phi)}{d\phi^2}$ evaluated at 
$\phi = \phi_{K}(x)$.
Using the three distinct kink solutions given by Eqs. (\ref{5}), (\ref{11})
and (\ref{16}) we now carry out the stability analysis in each of the three cases.

\subsection{Stability of Kink Solution $\phi^{I}_{K}(x)$}

On using the kink potential as given by Eq. (\ref{1}) and the first kink
solution as given by Eq. (\ref{5}) we find that 
\be\label{2.2}
V_K(x) = e^{2e^{-x}} e^{-2x} -3 e^{e^{-x}} e^{-x} [e^{-x} +1] + e^{-2x} 
+ 3 e^{-x} + 1\,, 
\ee
which is depicted in the inset of Fig. 3.  It may be noted that $V(\infty) = 1$ while $V(-\infty) = \infty$ so 
that the continuum begins at $\omega^2 = 1$. The corresponding kink zero mode is given by
\be\label{2.3}
\psi_{0}(x) = \frac{d\phi^{I}_{K}(x)}{dx} = e^{-e^{e^{-x}}}\, e^{e^{-x}}\,
e^{-x}\,.
\ee
The above zero mode is clearly nodeless and vanishes both as $x \rightarrow
\pm \infty$. Further, it is easy to check that the zero mode eigenfunction
(\ref{2.3}) satisfies the stability Eq. (\ref{2.1}) with the potential
$V_K(x)$ given by Eq. (\ref{2.2}) and with $\omega^2 = 0$. 

Summarizing, we find that indeed there is a gap between the zero mode and the
onset of the continuum in the case of the first kink solution.

\begin{figure}[h] 
\includegraphics[width= 5.1 in]{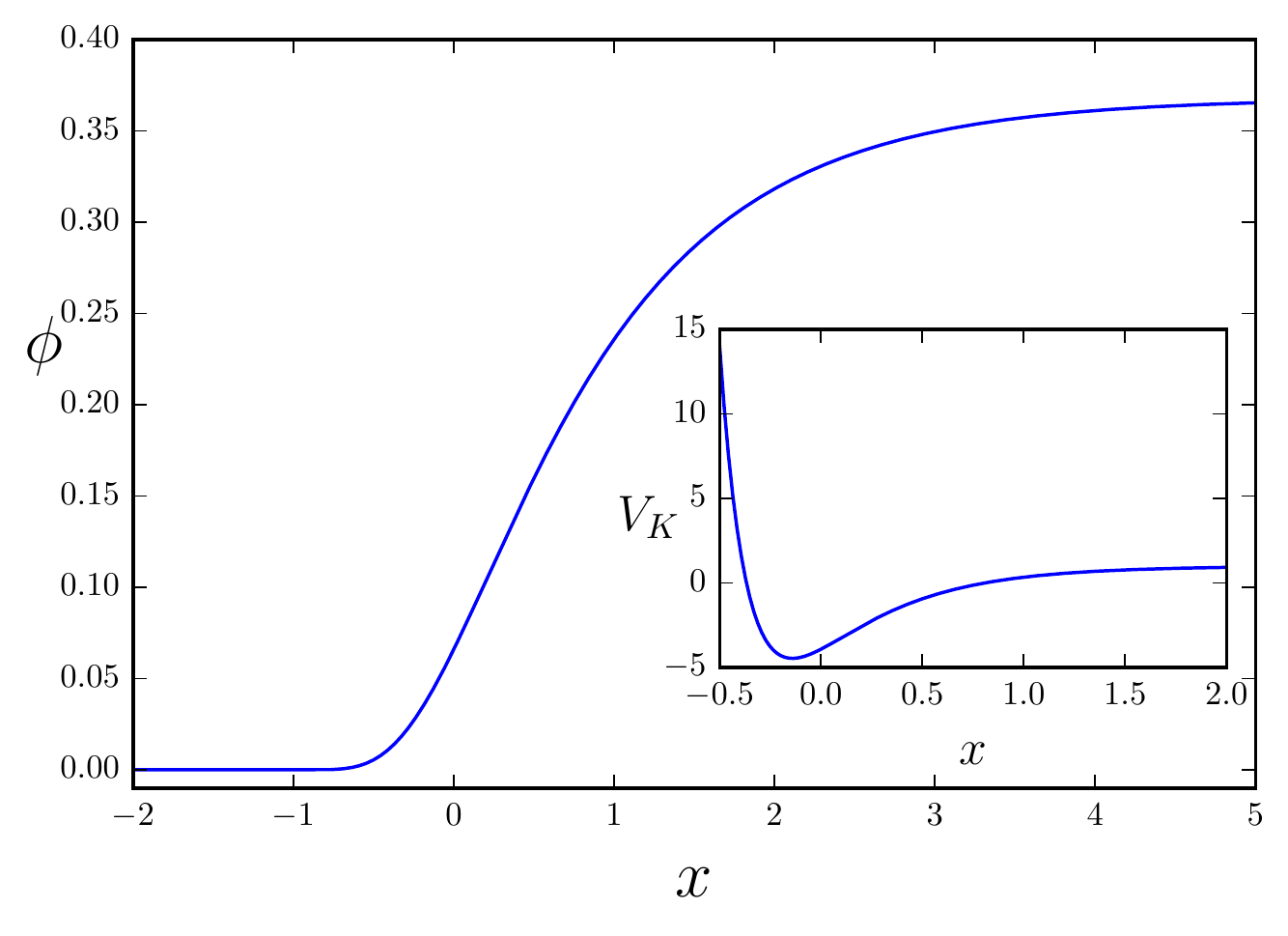}
\caption{A magnified version of the kink from $0\rightarrow 1/e$. Inset: The associated 
kink potential (see Eq. 23) which diverges super-exponentially for $x\rightarrow-\infty$ 
and asympotes to $1/e$ for $x\rightarrow\infty$. }

\end{figure}  

\subsection{Stability of Kink Solution $\phi^{II}_{K}(x)$}

On using the kink potential as given by Eq. (\ref{1}) and the second kink
solution as given by Eq. (\ref{11}) we find that 
\be\label{2.4}
V_K(x) = e^{-2e^{x}} e^{2x} -3 e^{-e^{x}} e^{x} [e^{x} -1] + e^{2x} 
- 3 e^{x} + 1\,, 
\ee
which is depicted in the inset of Fig. 4. It may be noted that $V(\infty) = \infty$ while $V(-\infty) = 1$ so 
that the continuum begins at $\omega^2 = 1$. The corresponding kink zero mode is given by
\be\label{2.5}
\psi_{0}(x) = \frac{d\phi^{II}_{K}(x)}{dx} = e^{-e^{-e^{x}}}\, e^{-e^{x}}\,
e^{x}\,.
\ee
The above zero mode is clearly nodeless and vanishes both as $x \rightarrow
\pm \infty$. Further, it is easy to check that the zero mode eigenfunction
(\ref{2.5}) satisfies the stability Eq. (\ref{2.1}) with the potential
$V_K(x)$ given by Eq. (\ref{2.4}) and with $\omega^2 = 0$. 

Summarizing, we find that indeed there is a gap between the zero mode and the
onset of the continuum in the case of the second kink solution.

\begin{figure}[h] 
\includegraphics[width= 5.1 in]{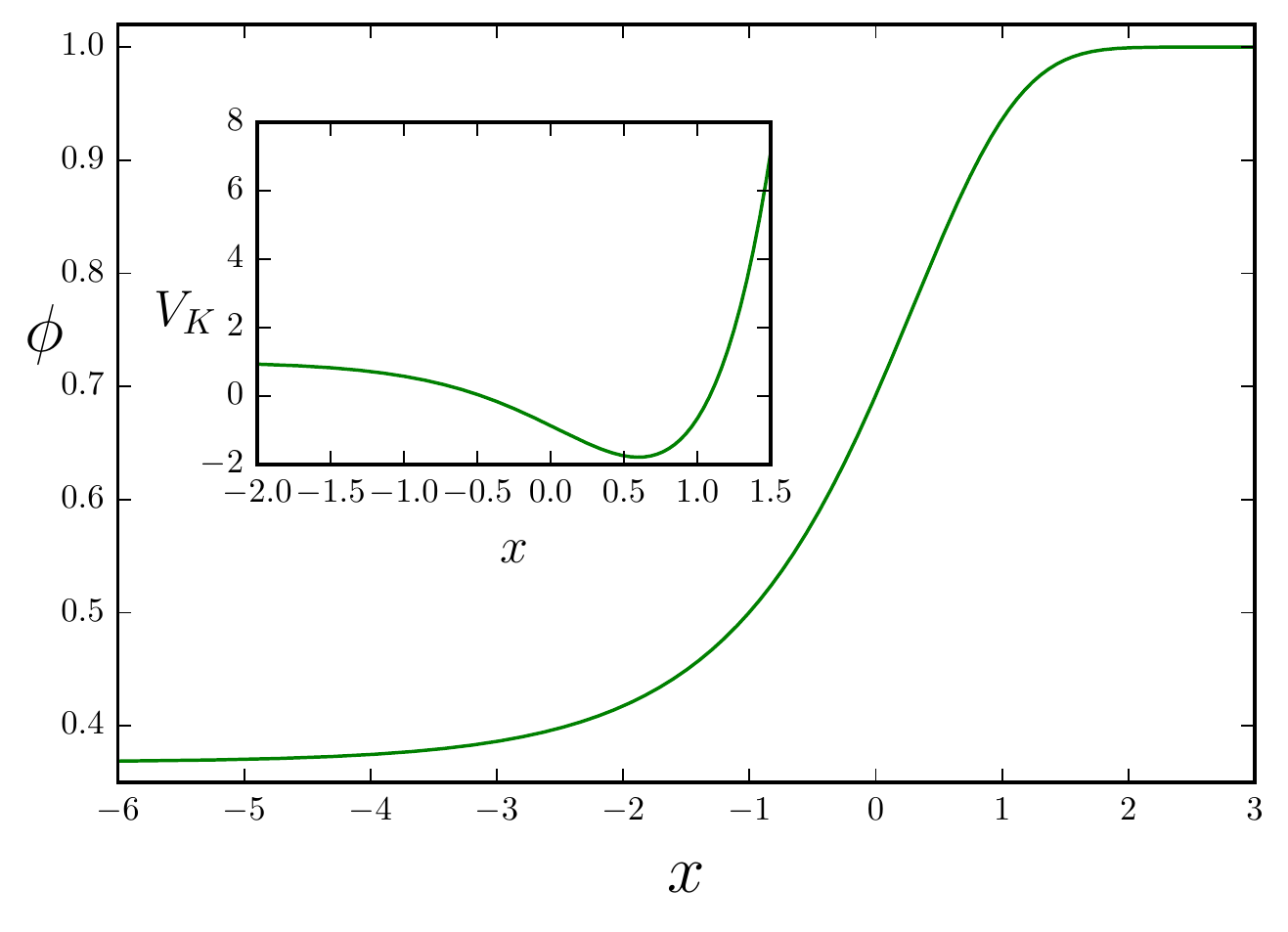}
\caption{A magnified version of the kink from $1/e\rightarrow 1$. Inset: The associated 
kink potential (see Eq. 25) which diverges exponentially for $x\rightarrow\infty$ and 
asymptotes to $1/e$ for $x\rightarrow-\infty$. }
\end{figure}  

\subsection{Stability of The Kink Solution $\phi^{III}_{K}(x)$}

On using the kink potential as given by Eq. (\ref{1}) and the third kink
solution as given by Eq. (\ref{16}) we find that 
\be\label{2.6}
V_K(x) = e^{-2e^{-x}} e^{-2x} +3 e^{-e^{-x}} e^{-x} [e^{-x} -1] + e^{-2x} 
- 3 e^{-x} + 1\,, 
\ee
which is depicted in the inset of Fig. 5. It may be noted that $V(\infty) = 1$ while $V(-\infty) = \infty$ so 
that the continuum begins at $\omega^2 = 1$. The corresponding kink zero mode is given by
\be\label{2.7}
\psi_{0}(x) = \frac{d\phi^{III}_{K}(x)}{dx} = e^{e^{-e^{-x}}}\, e^{-e^{-x}}\,
e^{-x}\,.
\ee
The above zero mode is clearly nodeless and vanishes both as $x \rightarrow
\pm \infty$. Further, it is easy to check that the zero mode eigenfunction
(\ref{2.7}) satisfies the stability Eq. (\ref{2.1}) with the potential
$V_K(x)$ given by Eq. (\ref{2.6}) and with $\omega^2 = 0$. 

We thus have seen that for all three kink solutions, the kink stability 
equation is such that there is a gap between the zero mode and the beginning 
of the continuum. 

\begin{figure}[h] 
\includegraphics[width= 5.1 in]{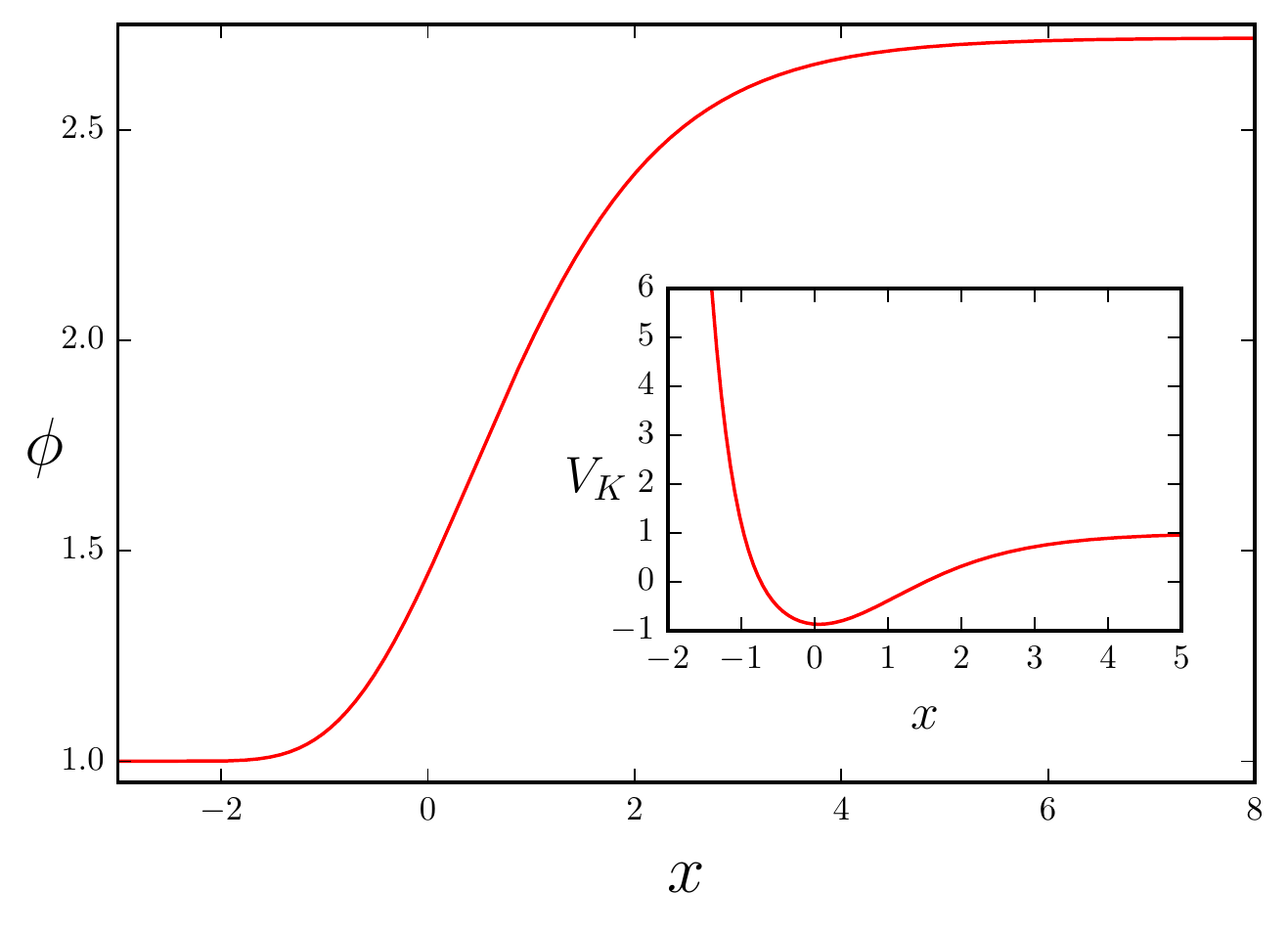}
\caption{A magnified version of the kink from $1\rightarrow e$. Inset: The associated 
kink potential (see Eq. 27) which diverges exponentially for $x\rightarrow-\infty$ 
and asympotes to $e$ for $x\rightarrow\infty$. }
\end{figure}  
\section{Kink-Kink and Kink-Antikink Interaction}

In this model we have three kinks, three mirror kinks and the corresponding
six antikinks. In particular, we have seen that while the kink tail around
$\phi = 0$ is super-super-exponential, the kink tails around $\phi = 1/e$
and $\phi = e$ are exponential. Finally the kink tail around $\phi = 1$ is
super-exponential. Using this information, we can immediately deduce the
nature of the KK and K-AK as well as AK-K interactions in various cases. 

To begin with, the interaction between the $(-1/e, 0)$ K and $(0, 1/e)$ K 
will be repulsive and 
super-super-exponential. On the other hand, the interaction between the 
$(1/e, 0)$ 
AK and $(0, 1/e)$ K will be attractive and super-super-exponental while the
interaction between the $(0, 1/e)$ K and $(1/e, 0)$ AK will be attractive and 
exponential.
On the other hand, the interaction between the $(0, 1/e)$ K and $(1/e, 1)$
K will be repulsive and exponential while the interaction  
between the $(1, 1/e)$ AK and $(1/e, 1)$ K, will be attractive 
and exponential.
Similarly, the interaction between the $(1/e, 1)$ K and $(1, 1/e)$ AK
will be super-exponential but attractive while the interaction between the 
$(1/e, 1)$ K and $(1, e)$ K will be repulsive but super-exponential. Likewise,
the interaction between the $(1/e, 1)$ AK and $(1, e)$ K will be attractive but
super-exponential. Finally, the interaction between the $(1, e)$ K and 
$(e, 1)$ AK will be attractive and exponential.    

As regards the kink-antikink sequences (on an infinite chain) in this model 
are concerned, there will be topological restrictions on the location of kinks 
and antikinks that are much more elaborate than those considered in the 
$\phi^6$ model for a first order transition \cite{sanati} and the $(\phi\ln\phi)^2$ 
potential for an infinite order transition \cite{pradeep}. 

\section{Kink Masses}

One can easily calculate the masses of all three kinks. As we show now, 
the formal expression for the kink mass is the same in all three cases, the 
only difference in the three cases comes from the different limits.
The kink mass is
given by
\be\label{4.1}
M_{K} = \int_{\phi_a}^{\phi_b}  d\phi\, \phi [(1/2) \ln(\phi^2)] 
\ln[(1/2)\ln(\phi^2)]\,,
\ee
where $\phi_a, \phi_b$ correspond to two contiguous minima between (see Fig. 1) 
which there is a kink solution. This integral is straightforward to evaluate, one
possible way is by using the
substitution $t = (1/2) \ln(\phi^2)$. We obtain
\be\label{4.2}
M_{K} = \frac{\phi^2}{4} [ \ln(\phi^2) - 1] \ln[(1/2)\ln(\phi^2)] 
- \frac{\phi^2}{4} +(1/4) Ei[\ln(\phi^2)]\,,
\ee
which is to be evaluated between the two limits $\phi_a$ and $\phi_b$.
Here Ei(x) is the exponential integral function \cite{erdelyi, grad}.  Let us now 
use appropriate limits and estimate the kink mass for all the three cases.

\vspace{0.2in} 
\noindent {\bf Mass of Kink I}

The kink-I goes from $\phi = 0$ to $\phi = 1/e$ as $x$ goes from $-\infty$
to +$\infty$. Hence its kink mass is obtained by evaluating the expression
for $M_{K}$ as given by Eq. (\ref{4.2}) between the limits $\phi_a = 0$ and 
$\phi_b = 1/e$. We find that
\be\label{4.3}
M^{I}_{K} = \frac{1}{4} [Ei(-2) - \frac{1}{e^2}]\,.  
\ee

\vspace{0.2in} 
\noindent{\bf Mass of Kink II}

The kink-II goes from $\phi = 1/e$ to $\phi = 1$ as $x$ goes from $-\infty$ to 
+$\infty$. Hence its kink mass is obtained by evaluating the expression 
(\ref{4.2}) between the limits $\phi_a = 1/e$ to $\phi_b = 1$. We find that 
\be\label{4.4}
M^{II}_{K} = -\frac{1}{4} [Ei(2) - \frac{1}{e^2} + 1]\,.
\ee

\vspace{0.2in} 
\noindent{\bf Mass of Kink III}

The kink-III goes from $\phi = 1/e$ to $\phi = 1$ as $x$ goes from $-\infty$ to 
+$\infty$. Hence its kink mass is obtained by evaluating the expression 
(\ref{4.2}) between the limits $\phi_a = 1$ to $\phi_b = e$. We find that 
\be\label{4.5}
M^{III}_{K} = \frac{1}{4} [Ei(2) - e^2 + 1]\,.
\ee

\section{Comparison with Higher Order Field Theories} 

It is instructive to compare the kink tails in the present case with other potentials that 
have seven minima, e.g. in the $\phi^{14}$ and even the higher order field theories ($\phi^{16}$ 
and $\phi^{18}$), briefly mentioned in \cite{KCS, Chapter} where the kinks have either 
a power law or an exponential tail. 

\subsection{Seven Minima of the $\phi^{14}$ Field Theory} 

Let us consider the following specific $\phi^{14}$ field theory model
\be\label{5.1}
V(\phi) = (1/2) \phi^2 (\phi^2 -1/e^2)^2 (\phi^2 - 1)^2 (\phi^2 - e^2)^2\,.
\ee
This model has 7 degenerate minima at $\phi = 0, \pm 1/e, \pm 1$ and at
$\pm e$ and hence 3 kink solutions and the corresponding three mirror
kink solutions as well as the corresponding six antikinks. The potential is 
depicted in Fig. 6 and also as a semilog plot in Fig. 7. 

It is worth noting that the values of the potential curvature at the seven degenerate minima are
\bea\label{5.2}
&&V''(0) = 1\,,~~ V''(\pm 1/e) = \frac{4(e^2-1)^4 (e^2+1)^2}{e^{12}}\,, 
\nonumber \\
&&V''(1) = \frac{4(e^2-1)^4}{e^4}\,,~~V''(\pm e) = 4(e^2-1)^4 (e^2+1)^2\,.
\eea

\begin{figure}[h] 
\includegraphics[width= 5.1 in]{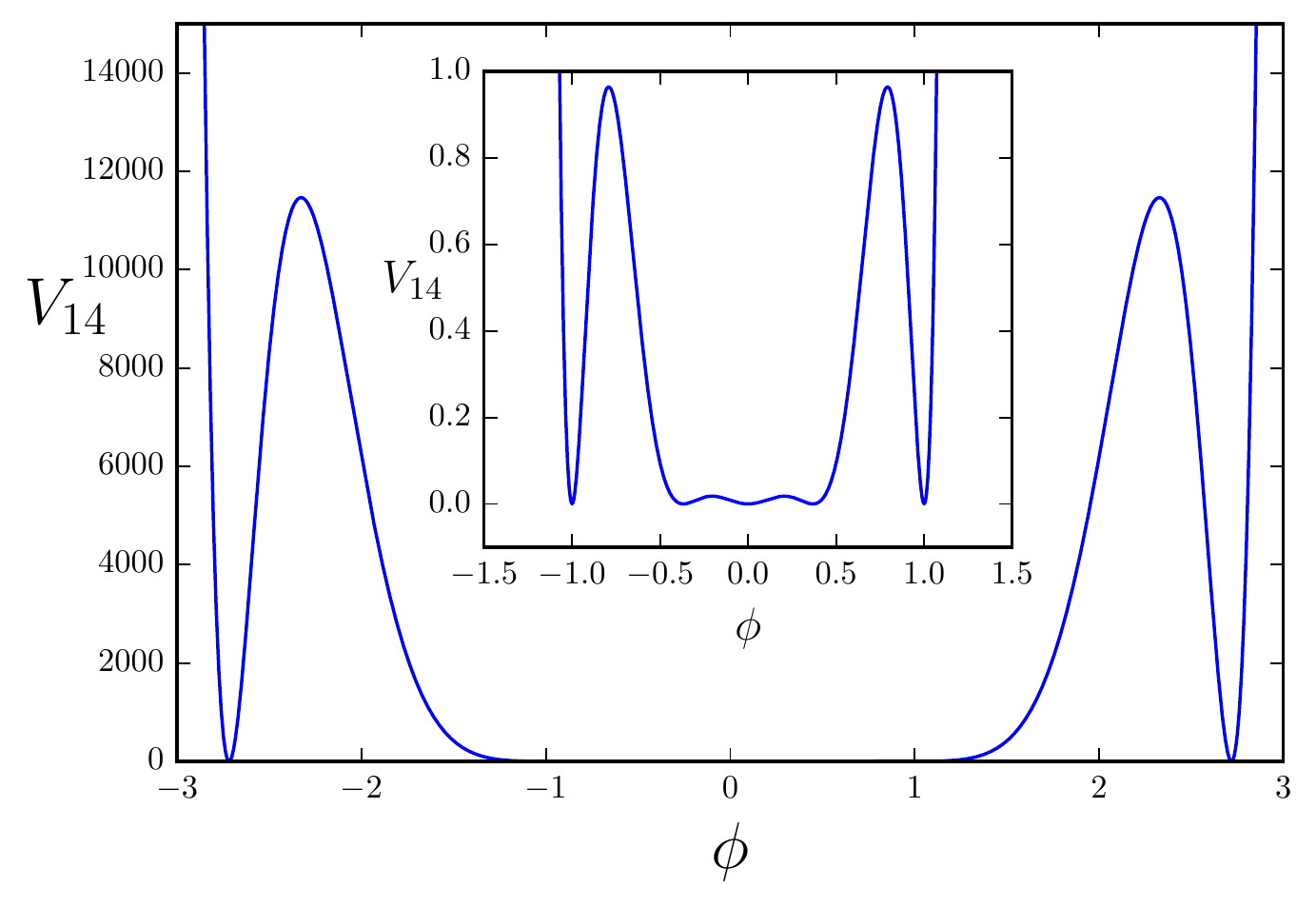}
\caption{The $\phi^{14}$ potential given by Eq. (33) with seven degenerate minima.  
Inset: Since the amplitudes of inner local maxima are so small compared to the outer 
maxima, a magnified version in the region $-1.5<\phi<1.5$ clearly shows the inner minima.  }
\end{figure} 

\begin{figure}[h] 
\includegraphics[width= 5.1 in]{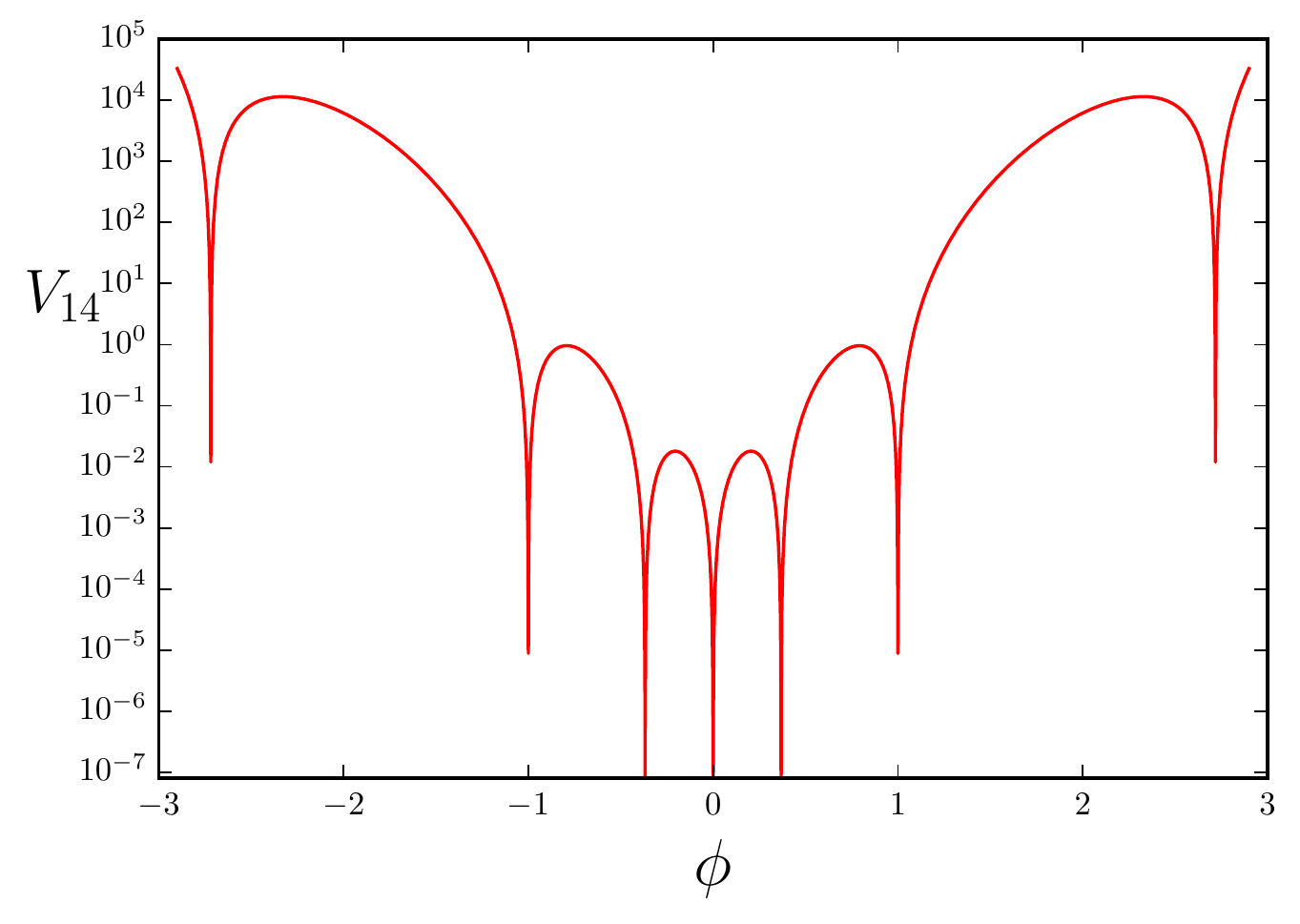}
\caption{The location of the seven minima of the $\phi^{14}$ potential can be easily 
seen in the semilog plot.  }
\end{figure} 

Let us now determine the three kink solutions, i.e. from $0$ to $1/e$, 
from $1/e$ to $1$ and from $1$ to $e$. The solutions for the corresponding 
three mirror kinks and the corresponding six antikinks can then be easily 
written down.
\vskip 0.1in 
\noindent{\bf Kink from $0$ to $1/e$}

In this case the self-dual first order equation is
\be\label{5.3}
\frac{d\phi}{dx} = \phi\, (1/e^2 - \phi^2) (1-\phi^2)\, (e^2 - \phi^2)\,.
\ee
Thus in this case
\be\label{5.4}
x = \int \frac{d\phi}{\phi\,(1/e^2 - \phi^2)\, (1-\phi^2)\,(e^2 - \phi^2)}\,.
\ee
Using partial fractions the integrand on the right hand side can be written as 
\be\label{5.5}
\frac{A_1}{\phi} + \frac{B_1 \phi}{1/e^2 - \phi^2} + \frac{C_1 \phi}{1-\phi^2}
+\frac{D_1 \phi}{e^2 - \phi^2}\,,
\ee
where
\be\label{5.6}
A_1 = 1\,,~~B_1 = \frac{e^6}{(e^2-1)^2 (e^2+1)}\,,~~C_1 
= -\frac{e^2}{(e^2-1)^2}\,,
~~D_1 = \frac{1}{(e^2-1)^2 (e^2+1)}\,.
\ee
This is easily integrated with the solution
\be\label{5.7}
x = \ln(\phi) -(B_1/2)\ln(1/e^2 -\phi^2) -(C_1/2) \ln(1-\phi^2) 
-(D_1/2)\ln(e^2 - \phi^2)\,.
\ee
Equation (40) is numerically inverted and the kink solution is depicted in Fig. 8. 
Thus, asymptotically
\be\label{5.8}
\lim_{x \rightarrow -\infty} \phi(x) = f_1(e) e^{x}\,,~~
\lim_{x \rightarrow \infty} \phi(x) = 1/e - g_1(e) e^{-2x/B_1}\,.
\ee
Here $f_1(e)$ and $g_1(e)$ are known constants.

\begin{figure}[h] 
\includegraphics[width= 5.1 in]{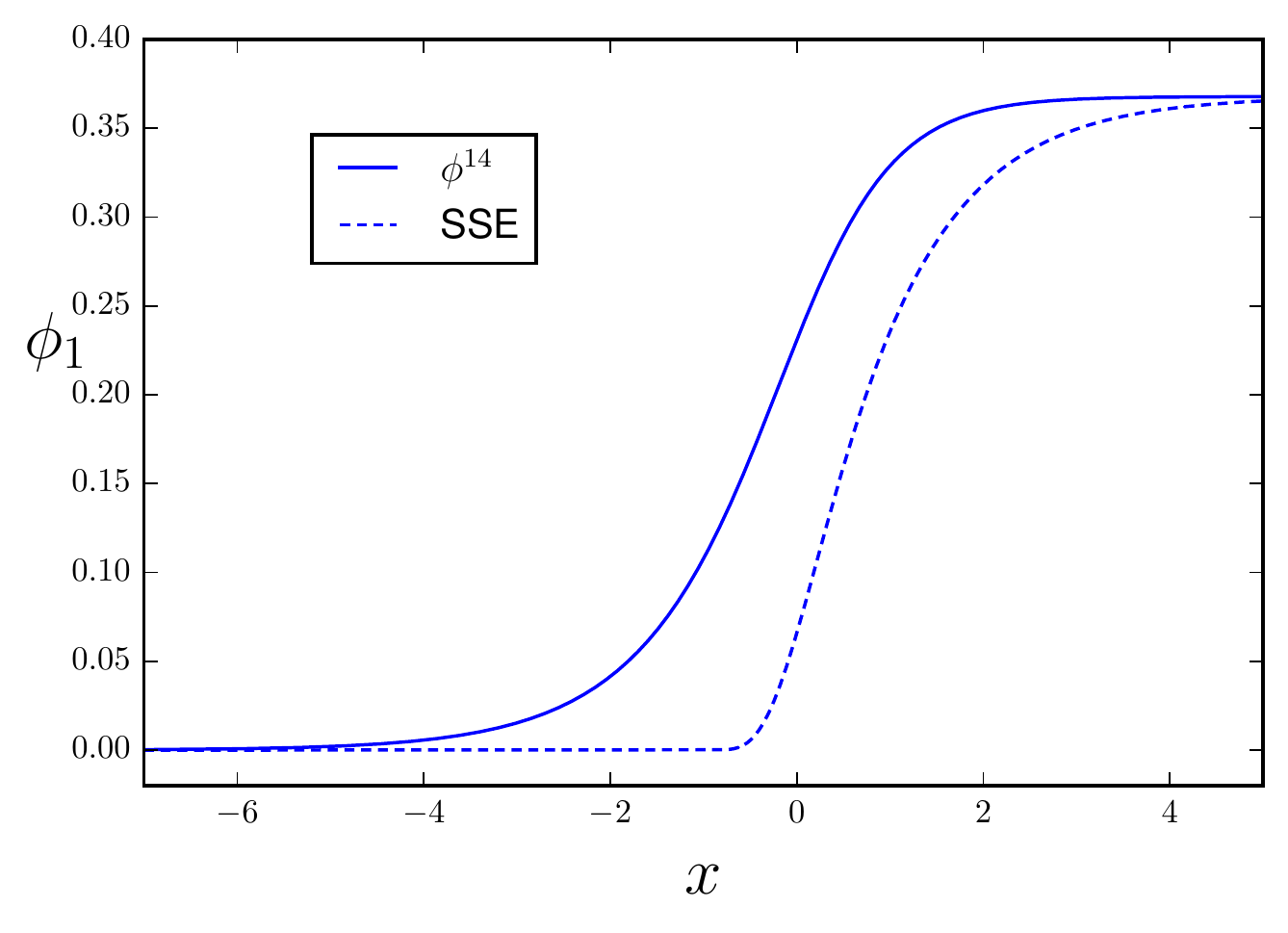}
\caption{Comparison of the $0\rightarrow 1/e$ kink with the corresponding 
super-super-exponetial (SSE) kink.  }
\end{figure}  

\begin{figure}[h] 
\includegraphics[width= 5.1 in]{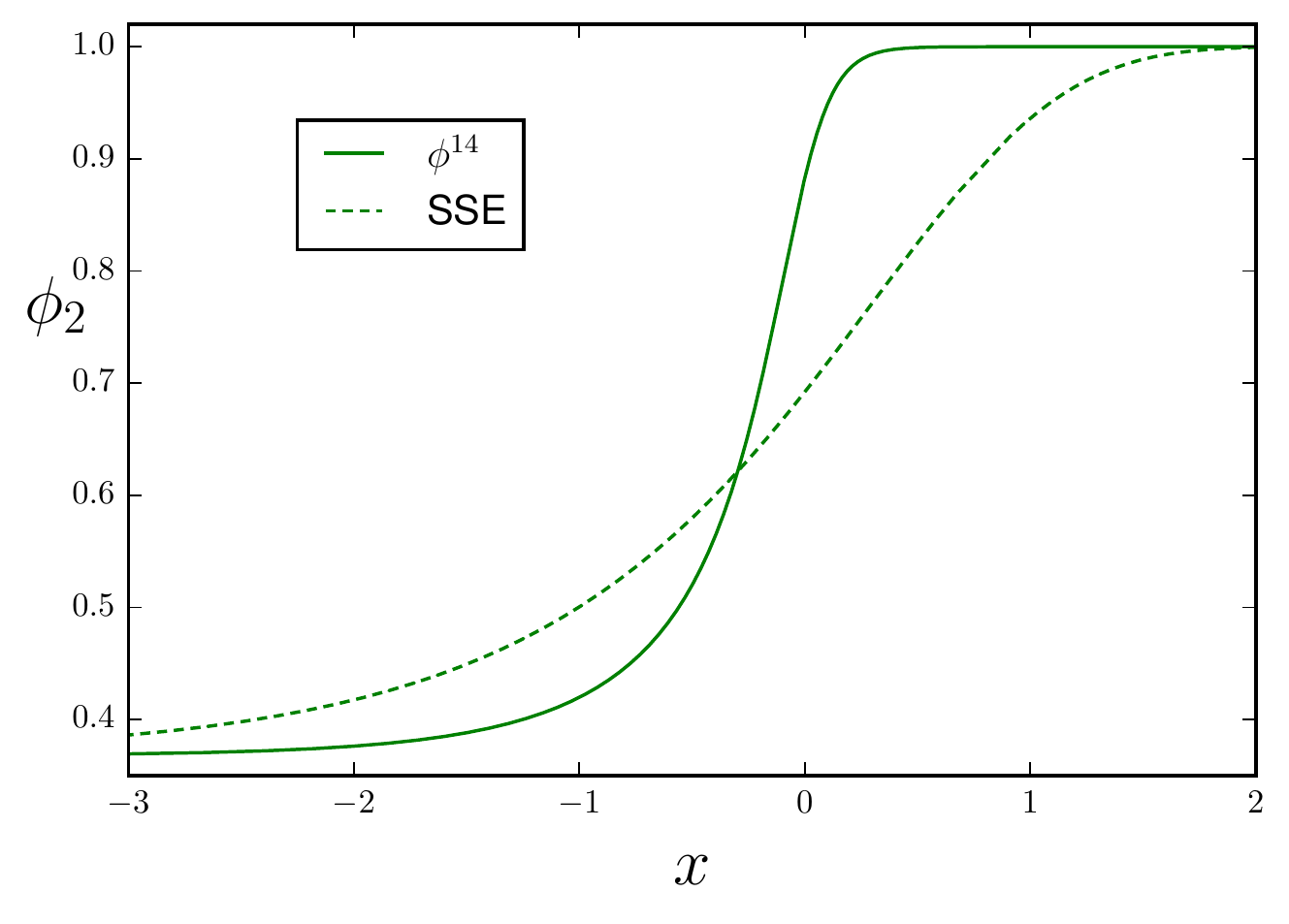}
\caption{Comparison of the $1/e\rightarrow 1$ kink with the corresponding 
super-super-exponetial (SSE) kink. }
\end{figure}  

\vskip 0.1in 
\noindent{\bf Kink from $1/e$ to $1$}

In this case the self-dual first order equation is
\be\label{5.9}
\frac{d\phi}{dx} = \phi\, (\phi^2- 1/e^2)\, (1-\phi^2)\, (e^2 - \phi^2)\,.
\ee
Thus in this case
\be\label{5.10}
x = \int \frac{d\phi}{\phi\,(\phi^2 -1/e^2)\, (1-\phi^2)\,(e^2 - \phi^2)}\,.
\ee
Again, using partial fractions the integrand on the right hand side can be written as 
\be\label{5.11}
\frac{A_2}{\phi} + \frac{B_2 \phi}{\phi^2 -1/e^2} + \frac{C_2 \phi}{1-\phi^2}
+\frac{D_2 \phi}{e^2 - \phi^2}\,,
\ee
where
\be\label{5.12}
A_2 = -1\,,~~B_2 = \frac{e^6}{(e^2-1)^2 (e^2+1)}\,,
~~C_2 = \frac{e^2}{(e^2-1)^2}\,,
~~D_2 = -\frac{1}{(e^2-1)^2 (e^2+1)}\,.
\ee
\begin{figure}[h] 
\includegraphics[width= 5.1 in]{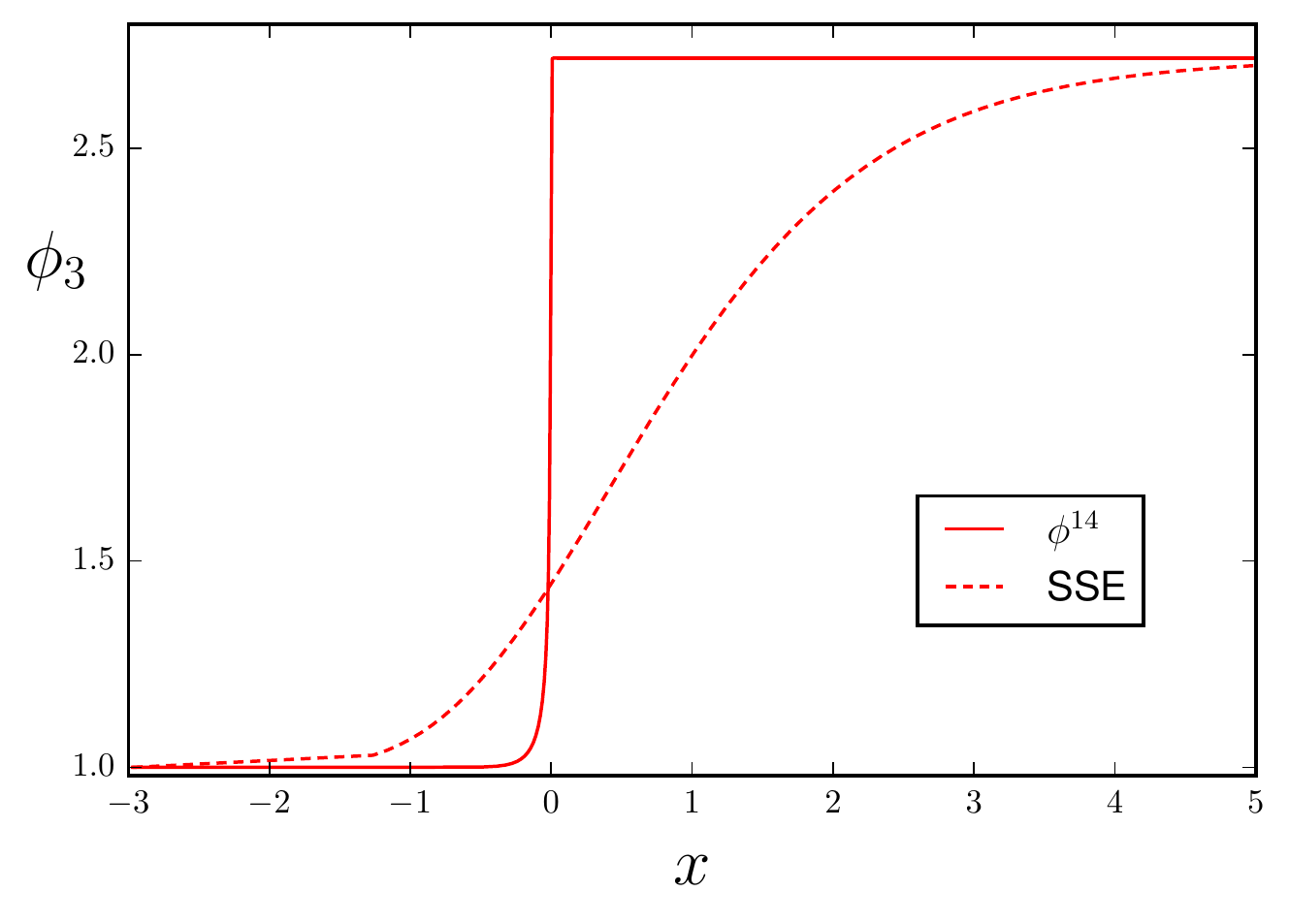}
\caption{Comparison of the $1\rightarrow e$ kink with the corresponding 
super-super-exponetial (SSE) kink. }
\end{figure} 

\begin{figure}[h] 
\includegraphics[width= 5.1 in]{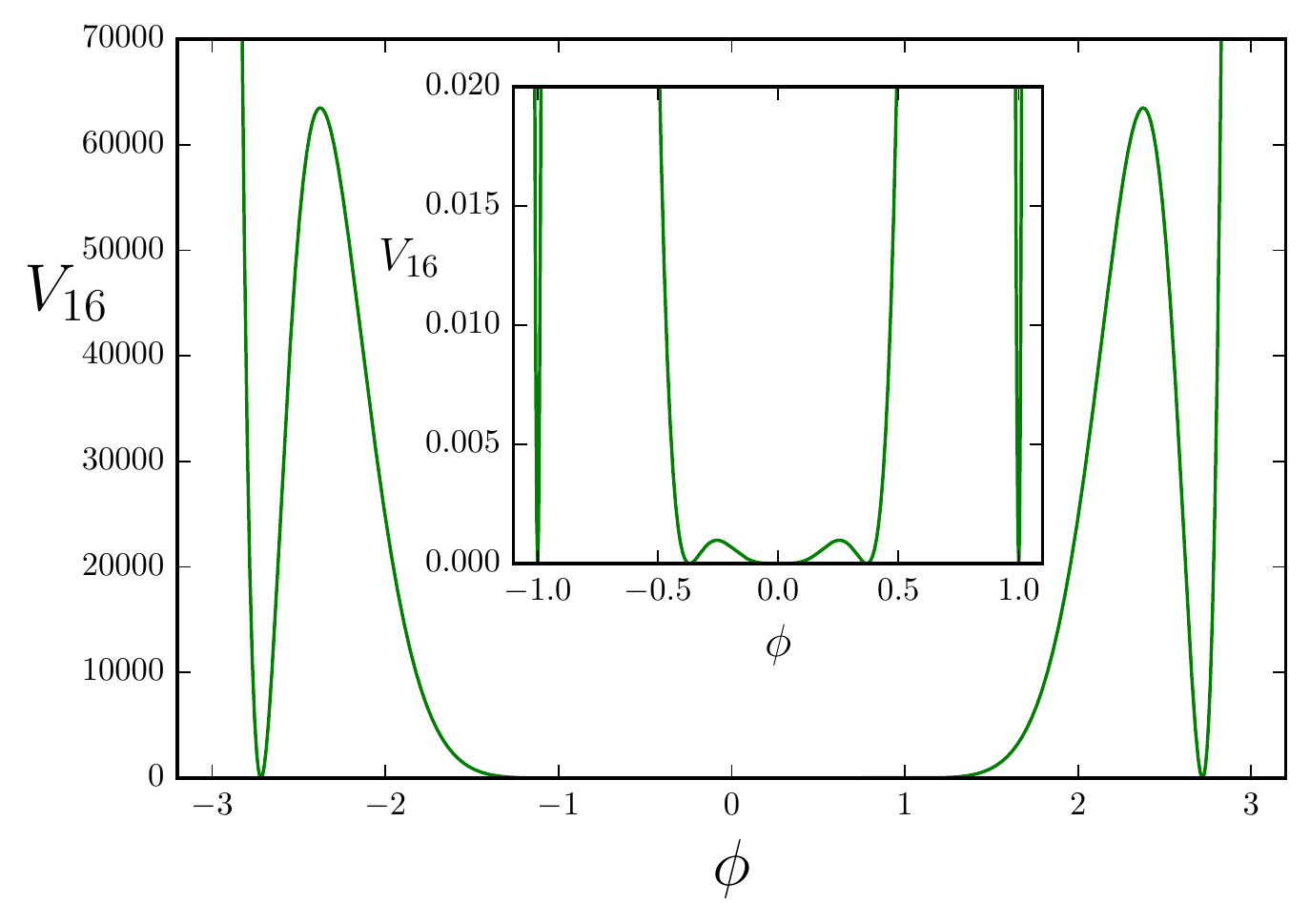}
\caption{The $\phi^{16}$ potential given by Eq. (54) with seven degenerate minima.  
Inset: Since the amplitudes of inner local maxima are so small compared to the outer 
maxima, a magnified version in the region $-1.1<\phi<1.1$ clearly shows the inner minima.  }
\end{figure}  
This is easily integrated with the solution
\be\label{5.13}
x = -\ln(\phi) +(B_2/2)\ln(\phi^2 -1/e^2) -(C_2/2) \ln(1-\phi^2) 
-(D_2/2)\ln(e^2 - \phi^2)\,.
\ee
Equation (46) is numerically inverted and the kink solution is depicted in Fig. 9. 
Thus, asymptotically
\be\label{5.14}
\lim_{x \rightarrow -\infty} \phi(x) = 1/e + f_2(e) e^{2x/B_2}\,,~~
\lim_{x \rightarrow \infty} \phi(x) = 1 - g_2(e) e^{-2x/C_2}\,.
\ee
Here $f_2(e)$ and $g_2(e)$ are known constants.

\begin{figure}[h] 
\includegraphics[width= 5.1 in]{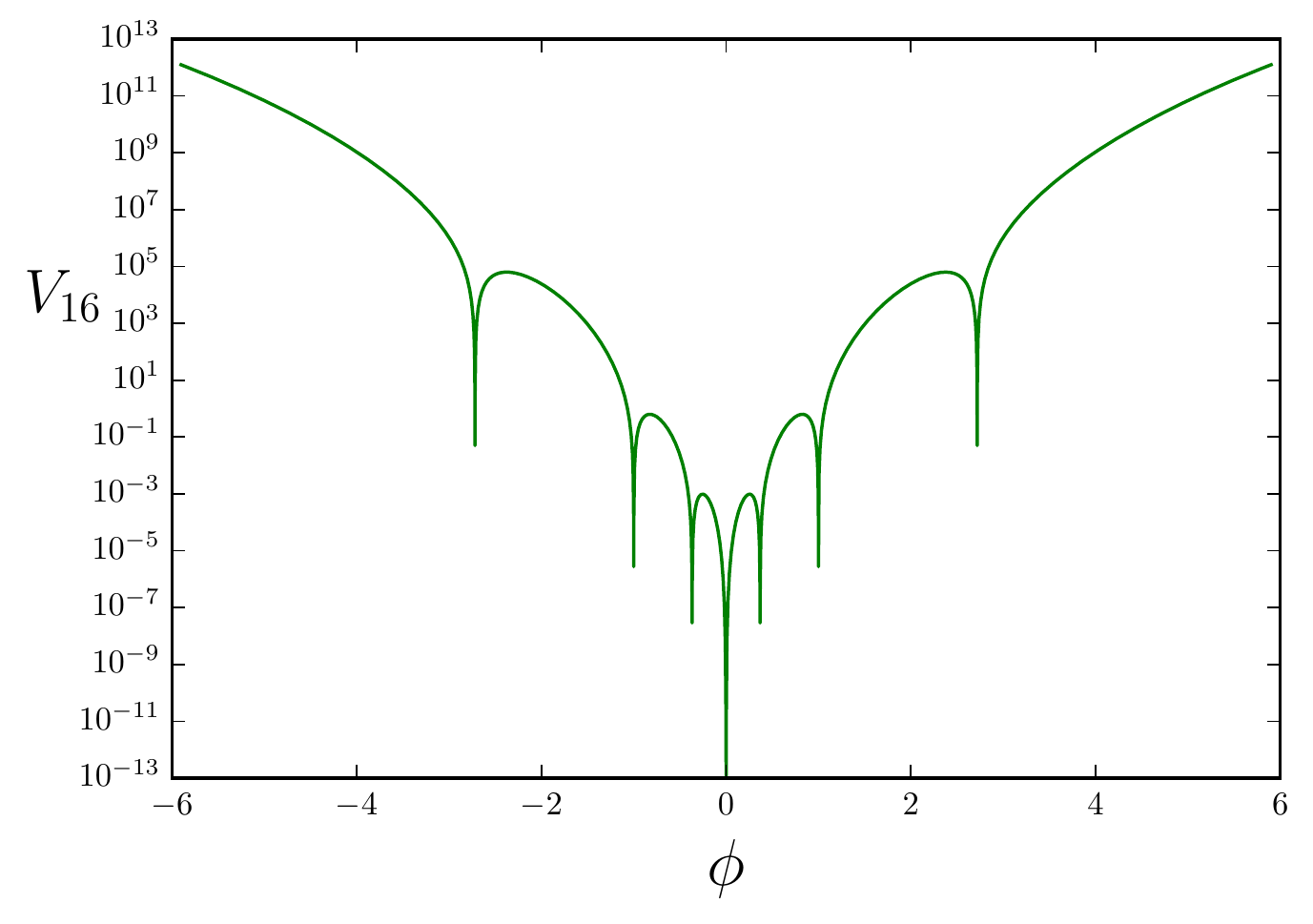}
\caption{The location of the seven minima of the $\phi^{16}$ potential can be easily 
seen in the semilog plot.  }
\end{figure} 

\begin{figure}[h] 
\includegraphics[width= 5.1 in]{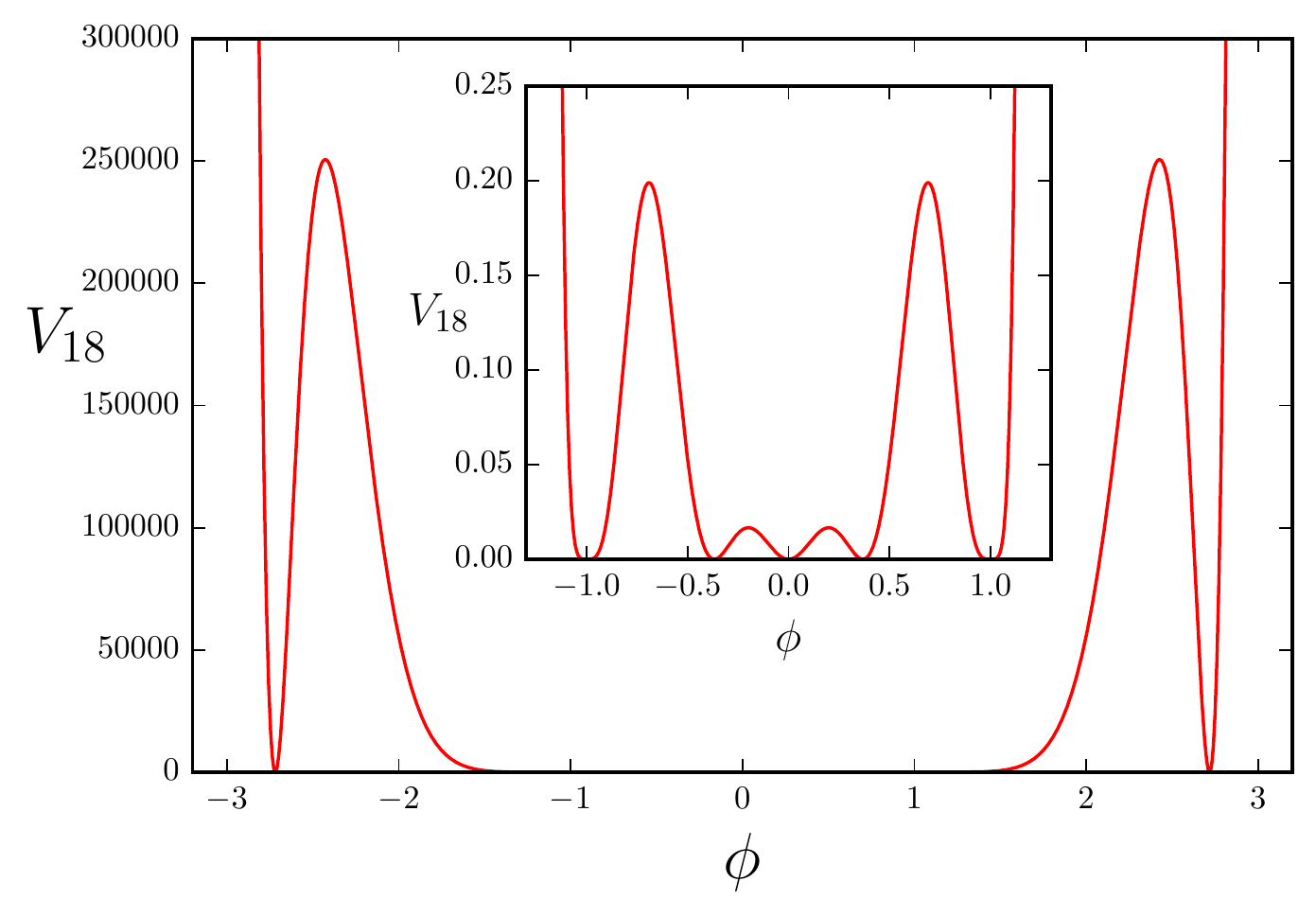}
\caption{The $\phi^{18}$ potential given by Eq. (55) with seven degenerate minima.  
Inset: Since the amplitudes of inner local maxima are so small compared to the outer 
maxima, a magnified version in the region $-1.2<\phi<1.2$ clearly shows the inner minima.  }
\end{figure} 

\begin{figure}[h] 
\includegraphics[width= 5.1 in]{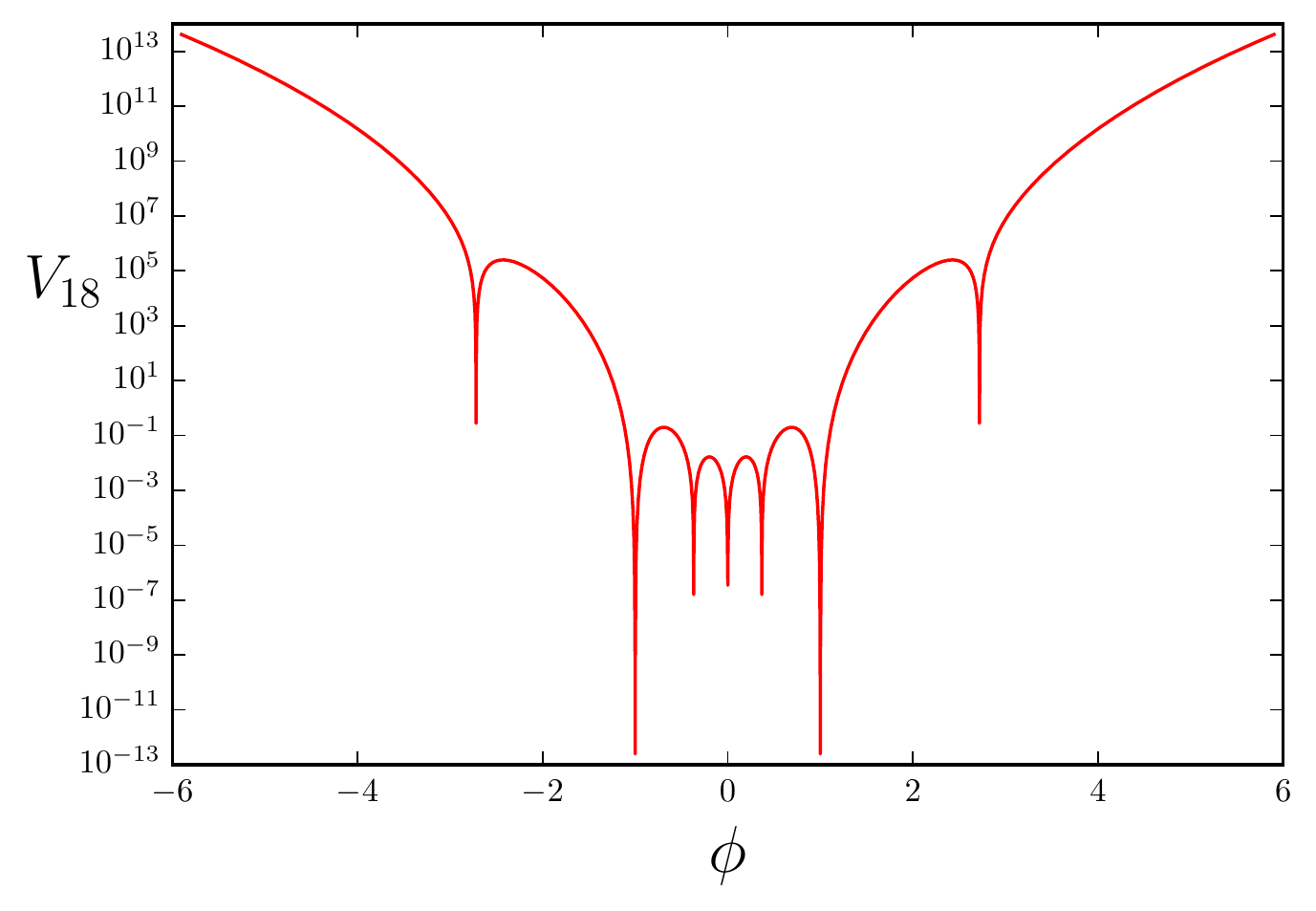}
\caption{The location of the seven minima of the $\phi^{18}$ potential can be easily 
seen in the semilog plot.  }
\end{figure}  

\vskip 0.1in 
\noindent{\bf Kink from $1$ to $e$}

In this case the self-dual first order equation is
\be\label{5.15}
\frac{d\phi}{dx} = \phi\, (\phi^2- 1/e^2)\, (\phi^2 -1)\, (e^2 - \phi^2)\,.
\ee
Thus in this case
\be\label{5.16}
x = \int \frac{d\phi}{\phi\,(\phi^2 -1/e^2)\, (\phi^2 -1)\,(e^2 - \phi^2)}\,.
\ee
Again, using partial fractions the integrand on the right hand side can be written as 
\be\label{5.17}
\frac{A_3}{\phi} + \frac{B_3 \phi}{\phi^2 -1/e^2} + \frac{C_3 \phi}{\phi^2 -1}
+\frac{D_3 \phi}{e^2 - \phi^2}\,,
\ee
where
\be\label{5.18}
A_3 = 1\,,~~B_3 = -\frac{e^6}{(e^2-1)^2 (e^2+1)}\,,
~~C_3 = \frac{e^2}{(e^2-1)^2}\,,
~~D_3 = \frac{1}{(e^2-1)^2 (e^2+1)}\,.
\ee
This is easily integrated with the solution
\be\label{5.19}
x = \ln(\phi) +(B_3/2)\ln(\phi^2 -1/e^2) +(C_3/2) \ln(\phi^2-1) 
-(D_3/2)\ln(e^2 - \phi^2)\,.
\ee
Equation (52) is numerically inverted and the kink solution is depicted in Fig. 10. 
Thus, asymptotically
\be\label{5.20}
\lim_{x \rightarrow -\infty} \phi(x) = 1 + f_3(e) e^{2x/C_3}\,,~~
\lim_{x \rightarrow \infty} \phi(x) = e - g_3(e) e^{-2x/D_3}\,.
\ee
Here $f_3(e)$ and $g_3(e)$ are known constants.

We can also obtain the kink stability potential for the three kinks (similar to the 
ones shown in the insets of Figs. 2 - 4) numerically but we do not pursue this here. 

\subsection{Seven Minima of the $\phi^{16}$ Field Theory} 

Consider a specific $\phi^{16}$ field theory model potential that is given by 
\be\label{5.21}
V(\phi) = (1/2) \phi^4 (\phi^2 -1/e^2)^2 (\phi^2 - 1)^2 (\phi^2 - e^2)^2\,.
\ee
This model also has 7 degenerate minima at $\phi = 0, \pm 1/e, \pm 1$ and at
$\pm e$ and hence 3 kink solutions and the corresponding three mirror
kink solutions as well as the corresponding six antikinks.  These kinks will have 
exponential tails except around $\phi=0$ which will be a power law tail.  The 
potential is depicted in Fig. 11 and also as a semilog plot in Fig. 12.  Using an  
analysis similar to the previous subsection we can obtain the kink solutions for  
the $\phi^{16}$ model as well but we do not depict them here. 

\subsection{Seven Minima of the $\phi^{18}$ Field Theory} 

The potential for a specific $\phi^{18}$ field theory model is given by 
\be\label{5.22}
V(\phi) = (1/2) \phi^2 (\phi^2 -1/e^2)^2 (\phi^2 - 1)^4 (\phi^2 - e^2)^2\,.
\ee

This model also has 7 degenerate minima at $\phi = 0, \pm 1/e, \pm 1$ and at
$\pm e$ and hence 3 kink solutions and the corresponding three mirror
kink solutions as well as the corresponding six antikinks.  These kinks will have 
exponential tails except around $\phi=1$ which will be a power law tail. The 
potential is depicted in Fig. 13 and also as a semilog plot in Fig. 14.  Using an  
analysis similar to subsection 6.1 we can also obtain the kink solutions for the 
$\phi^{18}$ model but we do not depict them here. 

\section{Conclusions and Some Open Problems}

Recently there has been a surge of interest in potentials harboring non-exponential 
kink tails \cite{KCS, Chapter, Gomes, Bazeia, Guerrero, Mello}, in particular power law, 
super-exponential \cite{pradeep} and power-tower \cite{powertower}.  Here we have 
introduced a class of potentials which have a kink solution with a super-super-exponential 
profile as well as a super-super-exponential tail.  According to the stability analysis of 
such kinks, there is a gap between the zero mode and the onset of continuum.  Since 
there are three different types of kinks between the seven minima there will be multiple 
topological restrictions \cite{pradeep} on the location of kinks/antikinks on an infinite 
chain that need to be elucidated. 

It would be desirable to have numerical studies of kink-kink collisions for kinks with 
super-super-exponential tails and to compare with collisions of other kinds of kinks: with 
exponential, super-exponential and power law tails. These potentials might find applications 
in the context of some unusual (such as infinite order) phase transitions or other physical 
contexts, e.g. multiple successive phase transitions \cite{KCS, Chapter}.  We might call the 
super-super-exponential profile as the ``super-Gumbel" distribution as opposed to the 
Gumbel distribution known in the area of extremal event statistics \cite{gumbel}. 

 \section{Acknowledgment}  We acknowledge very fruitful discussions with Ayhan Duzgun.  
 A. K. is grateful to INSA (Indian National Science Academy) for the award  of INSA 
 Senior Scientist position.  This work was supported in part by the U.S. Department of Energy.

\end{document}